# Surface Saturation Current Densities of Perovskite Thin Films from Suns-Photoluminescence Quantum Yield Measurements

Robert Lee Chin[1], Arman Mahboubi Soufiani[1], Paul Fassl[2], Jianghui Zheng[1,3], Eunyoung Choi[1], Anita Ho-Baillie[1,3], Ulrich Paetzold[2], Thorsten Trupke[1], Ziv Hameiri[1]

[1] The University of New South Wales, Sydney, Australia

[2] Karlsruhe Institute of Technology (KIT), Karlsruhe, Germany

[3] The University of Sydney, Sydney, Australia

Email address: r.leechin@unsw.edu.au

## Abstract

We present a simple, yet powerful analysis of Suns-photoluminescence quantum yield measurements that can be used to determine the surface saturation current densities of thin film semiconductors. We apply the method to state-of-the-art polycrystalline perovskite thin films of varying absorber thickness. We show that the non-radiative bimolecular recombination in these samples originates from the surfaces. To the best of our knowledge, this is the first study to demonstrate and quantify non-linear (bimolecular) surface recombination in perovskite thin films.

## 1. Introduction

The photoluminescence quantum yield (PLQY) is an important optoelectronics measurement that can determine the implied photovoltaic quality of semiconductors [1], [2]. PLQY measures the number of emitted photoluminescence (PL) photons, relative to the number of absorbed photons [3]. Often, the PLQY is measured using a spectrometer [1], such that the spectral photon flux is obtained. When PLQY measurements are performed as a function of the light intensity or Suns (often named 'Suns-PLQY'), the excess carrier density, $\Delta n$, is varied providing information about the injection-dependence of the PLQY [1].

For lead-based perovskite thin films (PTFs), PLQY measurements have been used to quantify the implied open-circuit voltage, $iV_{OC}$, and the $iV_{OC}$ deficit, $\Delta iV_{OC}$ [1], [4]–[6]. The $iV_{OC}$ represents the quasi-Fermi level splitting (QFLS), or the maximum possible open-circuit voltage that could be obtained if the film were converted into a solar cell [7], while $\Delta iV_{OC}$ represents the loss in $iV_{OC}$ due to defect-mediated or non-radiative recombination [8]. Suns-PLQY measurements have been used to quantify the non-radiative recombination of PTFs via extraction of the implied efficiency potential and ideality factor, $n_{id}$ [1], [9], [10].

One of the main limitations of state-of-the-art PTFs is non-radiative bimolecular recombination (NRBR) [11]. NRBR refers to recombination which has the same injection-dependence or $n_{id}$ ($n_{id}$ = 1) as radiative bimolecular recombination but is mediated by defects, and hence does not produce photons (as expected in radiative recombination) [12]. For context, the radiative recombination mechanism is described as "bimolecular" because each recombination event involves the recombination of a free electron with a free hole [13]. NRBR in PTFs has been studied for more than half a decade, mostly using PLQY and/or time-resolved PL decay (TR-PL) measurements [11], [12], [14]–[18]. However, the exact nature and spatial origin of NRBR remains elusive. Previous suggestions included trap-Auger recombination inside the bulk [12], [19], yet first principles calculations from Staub *et al.* demonstrated that the trap-Auger mechanism is unlikely as the required bulk defect densities, $N_t$, is much larger than the experimentally determined value (above $10^{17}$ cm$^{-3}$ compared to below $10^{16}$ cm$^{-3}$) [16]. Brenes *et al.* subjected PTFs to light and atmospheric treatments and observed an increase in the PLQY and a concurrent decrease in the NRBR as determined from TR-PL [18]. They proposed that the treatments passivate surface states thereby reducing NRBR, however, no supportive evidence has been provided.

In this study, we use Suns-PLQY measurements to quantify and determine the spatial origin of NRBR in state-of-the-art PTFs. Our analysis involves converting Suns-PLQY data to dark implied current density-voltage ($iJ$-$V$) curves and subsequent extraction of the recombination current parameter, $J_0$, which we use to quantify NRBR. To validate the results, PTFs of the same composition but varying absorber thickness, $W$, were used.

## 2. Experimental Methods

To prepare the perovskite $Cs_{0.05}FA_{0.79}MA_{0.16}Pb(I_{0.83}Br_{0.17})_3$ precursor solution, cesium iodide (CsI, Sigma-Aldrich), formamidinium iodide (FAI, GreatCell Solar Materials), lead iodide (PbI$_2$, TCI), methylammonium bromide (MABr, GreatCell Solar Materials), and lead bromide (PbBr$_2$, Sigma-Aldrich) were stoichiometrically weighed and dissolved in a mixed solvent of N,N-dimethylformamide (DMF, TCI) and dimethyl sulfoxide (DMSO, Alfa Aesar) (4:1 v/v) with different concentrations: 1.1 M (thick perovskite layer), 0.74 M (intermediate layer), and 0.5

M (thin layer). Glass substrates were sequentially cleaned with detergent, deionised water, acetone, and isopropanol. After that, the cleaned glass substrates were treated with ultraviolet ozone (UVO) for 15 mins and then transferred into a nitrogen-filled glovebox. The precursor solution was spin-coated on the substrates at 2,000 rpm for 20 s, followed by 6,000 rpm for 30 s. During the spin-coating, 100 µl chlorobenzene (Sigma-Aldrich) was quickly dispensed 5 s prior to the end of the spin-coating process. The films were then annealed at 100 °C for 10 min on a hot plate, producing a dense perovskite film as indicated by secondary electron microscopy (SEM) images. For brevity, this PTF composition is denoted "Br17".

Suns-PLQY measurements were carried out inside an integrating sphere (15 cm diameter, LabSphere), flushed with nitrogen gas. A green laser (LD-515-10MG, Roithner Lasertechnik) with a full-width-half-maximum (FWHM) spot size of ~778 µm was directed into the sphere via a small entrance port. An optical fiber collected the emission from the exit port of the sphere and guided it to the spectrometers (AvaSpec-2048x64TEC, Avantes). The spectral response was calibrated using a calibration lamp (HL-3plus-INT-Cal, Ocean Insight), giving a relative uncertainty of about ±3% in the spectral range of the PL emission (1.24 eV to 2.06 eV or 600 nm to 1,000 nm). Raw measured spectra were converted to power spectra by normalising for integration time. The integration time was 0.2 s for most of the measurements. The PLQY was determined using the method described by de Mello *et al*. [3]. The samples were placed at an angle of 15° with respect to the laser beam to avoid specular reflectance toward the entrance port. To stabilise both the spectral PL magnitude and shape, the samples were light-soaked at an intensity of ~30 Suns for 3 min, after which the Suns-PLQY measurements were performed.

Spectroscopy ellipsometry (SE) was used to determine the spectral absorptivity and thickness of the PTFs. An ellipsometer (M2000, J. A. Woollam) was used to measure the complex refractive index of the PTF on glass in the wavelength range of 210 to 1,000 nm. The ratio of the change in the polarisation of the light reflected from the sample was measured in ambient air at incident angles of 55°, 65°, and 75° to establish a model for the room temperature optical constants of the Br17 PTF. For this purpose, the computer software WVASEVR [ref] was used. To add an extra constraint on the ellipsometry fits, the experimental transmission spectrum of the measured samples, taken using a spectrophotometer (Lambda 1050, Perkin Elmer), was simultaneously fitted alongside the SE data. The curve fits are presented in S1 of the Supplementary Information (SI). The transfer matrix method (TMM) is used to calculate the PTF spectral absorptivity [20]. The thicknesses determined from the spectroscopy ellipsometry and transmission (SE/T) are summarised in **Table 1**.

**Table 1** Thicknesses of the PTF samples used in this study, as determined by SE/T.

|  | Thick | Intermediate | Thin |
|---|---|---|---|
| $W$ [nm] | 469.3 ± 4.0 | 262.2 ± 2.0 | 159.5 ± 5.5 |

## 3. Analysis Methods

The PLQY is defined as the ratio of emitted PL photons to the absorbed excitation photons:

$$\text{PLQY}(J_{\text{rec}}) = \frac{\int \phi_{\text{PL}}(\hbar\omega, J_{\text{rec}})\, d\hbar\omega}{\phi_{\text{ex}}} \qquad (1)$$

where $\phi_{PL}(\hbar\omega, J_{rec})$ is the *emitted* absolute spectral PL photon flux (cm$^{-2}$·s$^{-1}$·eV$^{-1}$) and $\phi_{ex}$ is the *absorbed* excitation photon flux (cm$^{-2}$·s$^{-1}$). $\phi_{PL}$ is a function of the photon energy, $\hbar\omega$, and is specified at a particular value of $\phi_{ex}$. The $J_{rec}$, represents the current density due to the net generation (recombination) rate of excess charge carriers and is given by: $J_{rec} = q \cdot \phi_{ex}$. It is noted that PTFs exhibit a significant fraction (~50%) of emission from the glass edges due to radiative recombination photons scattering from the non-planar PTF surface and waveguiding into the glass [21]. Therefore, this analysis requires that Suns-PLQY measurements be performed inside a 4π integrating sphere [3].

The Lasher-Stern Würfel (LSW) equation [7], [22] relates $\phi_{PL}(\hbar\omega)$ to the *iV*$_{OC}$:

$$\phi_{PL}(\hbar\omega, J_{rec}, T) = \text{Abs}(\hbar\omega) \cdot \phi_{BB}(\hbar\omega, T) \cdot \exp\left[\frac{iV_{OC}(J_{rec})}{k_B T}\right] \quad (2)$$

Abs($\hbar\omega$) is the spectral band-to-band absorptivity, $\phi_{BB}(\hbar\omega, T)$ is the Blackbody spectral photon flux emitted into the full sphere (4π sr) at carrier temperature *T*, and $k_B T$ is the thermal energy. Equation (2) may be linearised to extract *iV*$_{OC}$ [23]–[25]:

$$\ln\left[\frac{2\pi^2 \hbar^3 c_0^2}{(\hbar\omega)^2} \cdot \frac{\phi_{PL}(\hbar\omega,)}{\text{Abs}(\hbar\omega)}\right] = \frac{1}{k_B T}[\hbar\omega - iV_{OC}(J_{rec})] \quad (3)$$

Ideally, plotting the left-hand-side of Equation (3) as a function of $\hbar\omega$ forms a straight line with a slope of $1/k_B T$ and a y-intercept of -*iV*$_{OC}/k_B T$.

Subsequently, the radiative-limited implied voltage, *iV*$_{OC,rad}$, can be determined from *iV*$_{OC}$ and PLQY [8]:

$$iV_{OC,rad}(J_{rec}) = iV_{OC}(J_{rec}) - \underbrace{k_B T \cdot \ln[\text{PLQY}(J_{rec})]}_{\Delta iV_{OC}} \quad (4)$$

*iV*$_{OC,rad}$ represents *iV*$_{OC}$ when no non-radiative recombination is present. $J_{rec}$ plotted as a function of *iV*$_{OC}$ represents the dark *iJ-V* curve of the PTF: $J_{rec}(iV_{OC}) = iJ(iV)$, where $iJ(iV)$ is the implied dark current-density as a function of the implied voltage. The *iJ-V* curve can be decomposed into the *iJ-V* curves due to radiative (*iJ*$_{rad}$) and non-radiative (*iJ*$_{nr}$) recombination, respectively:

$$iJ(iV) = iJ_{rad}(iV) + iJ_{nr}(iV) \quad (5)$$

$J_{rec}$ plotted as a function of *iV*$_{OC,rad}$ represents the radiative dark *iJ-V* curve: $J_{rec}(iV_{OC,rad}) = iJ_{rad}(iV)$. As the radiative recombination has a $n_{id}$ of unity, the radiative dark *iJ-V* curve can be modelled as:

$$iJ_{rad}(iV) = J_{0,rad} \cdot \left(\exp\left[\frac{iV}{k_B T}\right] - 1\right) \quad (6)$$

where $J_{0,rad}$ is the radiative recombination parameter (A·cm$^{-2}$) [26]. If NRBR is the dominant non-radiative recombination mechanism, $iJ_{nr}(iV)$ is:

$$iJ_{nr}(iV) = J_{0,nr} \cdot \left(\exp\left[\frac{iV}{k_B T}\right] - 1\right) \quad (7)$$

where $J_{0,nr}$ is the recombination parameter associated with NRBR (A·cm$^{-2}$). Combining Equations (5) and (6), $iJ(iV)$ can be expressed as:

$$iJ(iV) = (J_{0,\text{rad}} + J_{0,\text{nr}}) \cdot \left( \exp\left[\frac{iV}{k_\text{B}T}\right] - 1 \right) \qquad (8)$$

By varying $W$, one can determine the spatial origin of $J_{0,nr}$. Accounting for NRBR at the surfaces and inside the bulk [26], $J_{0,nr}$ can be expressed as:

$$J_{0,\text{nr}} = J_{0,\text{s}} + \underbrace{q \cdot W \cdot n_\text{i}^2 \cdot B_{\text{bulk,nr}}}_{J_{0,\text{bulk,nr}}} + \qquad (9)$$

where $J_{0,s}$ and $J_{0,bulk,nr}$ are the non-radiative saturation current densities for surface and bulk NRBR, respectively. $B_{bulk,nr}$ is the NRBR coefficient [11], [12], [15], [18], and $n_i$ is the intrinsic carrier density [27].

## 4. Results and Discussion

Unless otherwise stated, all the presented measurement results are for the intermediate PTF. **Figure 1a** illustrates the Suns-PLQY measurements, represented as $\phi_{PL}(\hbar\omega)$, which cover nearly three orders of magnitude of $\phi_{ex}$. Throughout the wide range of Suns, we observed no changes in the shape of $\phi_{PL}(\hbar\omega)$, indicating the reasonably stable composition of the studied PTFs. Additionally, the absence of low-energy shoulders in $\phi_{PL}(\hbar\omega)$ suggests that phase segregation [28] does not have a significant macroscopic impact on the Suns-PLQY measurements of the investigated samples.

**Figure 1b** shows PLQY as a function of $\phi_{ex}$ in Suns, where 1-Sun is defined as equivalent to a photogenerated current density $J_L$, of 18.7 mA.cm$^{-2}$ (see Section S2 of the SI). We observe the PLQY remains consistently within (25.2 ± 3.3) % over almost two orders of magnitude ($\phi_{ex}$ > 0.2 Suns). The PLQY may be defined additionally to Equation (1) as the external radiative recombination rate ($R_{rad,ext}$) relative to the net recombination rate ($R_{tot}$): PLQY = $R_{rad,ext}/R_{tot}$ [29]. A plateau of the PLQY indicates that the dominant recombination mechanism has the same injection dependence as the radiative recombination. The difference between this plateau value and unity is attributed to non-radiative recombination. Thus, NRBR can be identified in Suns-PLQY as a plateau of the PLQY (PLQY < 1) with changing Suns. This constancy suggests that the dominant non-radiative recombination mechanism is indeed NRBR.

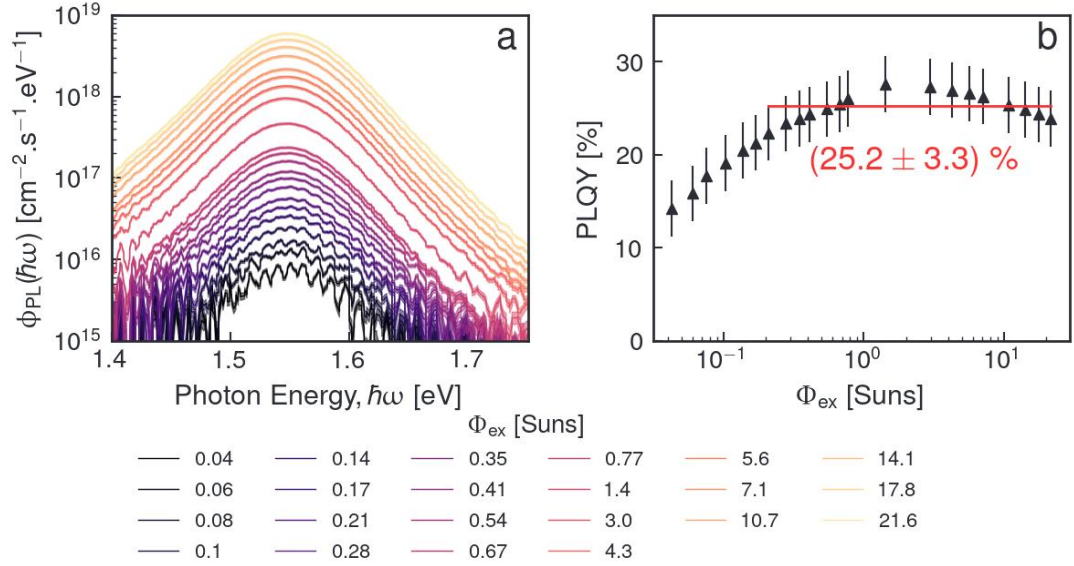

**Figure 1** (a) $\phi_{PL}(\hbar\omega)$, colour coded to $\phi_{ex}$. The shaded regions represent the measurement uncertainty due to the absolute calibration (±3%) and spectrometer dark noise. (B) PLQY vs. $\phi_{ex}$. The red line shows the approximate plateau value of the PLQY, (25.2 ± 3.3) %, indicating the presence of NRBR.

**Figure 2a** demonstrates the application of Equation (3) to the 1-Sun $\phi_{PL}(\hbar\omega)$ of the intermediate PTF (see also Section S2 of SI). The fitting region was carefully selected to ensure a high signal-to-noise ratio (SNR) and avoid the "edge artefact" caused by low-energy photons scattered at the film surfaces and guided towards the glass edges [21]. This fitting region (1.57 eV < $\hbar\omega$ < 1.67 eV), is indicated by the pair of black, dashed vertical lines. The fit quality appears to be good, reflected by the fitted $iV_{OC}$ and $T$ uncertainties of only ±1 mV and ±0.7 K, respectively. We note that the fitting range from above determines a carrier temperature close to the room temperature of about (296.7 ± 2.5) K [(22.5 ± 2.5)°C]. We provide an analysis of the fitting sensitivity in Section S3 of the SI.

**Figure 2b** displays the back-calculated spectral absorptivity (red line), which exhibits good agreement with the measured spectral absorptivity (black pentagons) in the region unaffected by the edge artefact ($\hbar\omega$ > 1.57 eV). This procedure is repeated for each $\phi_{PL}(\hbar\omega)$ curve to determine $iV_{OC}$.

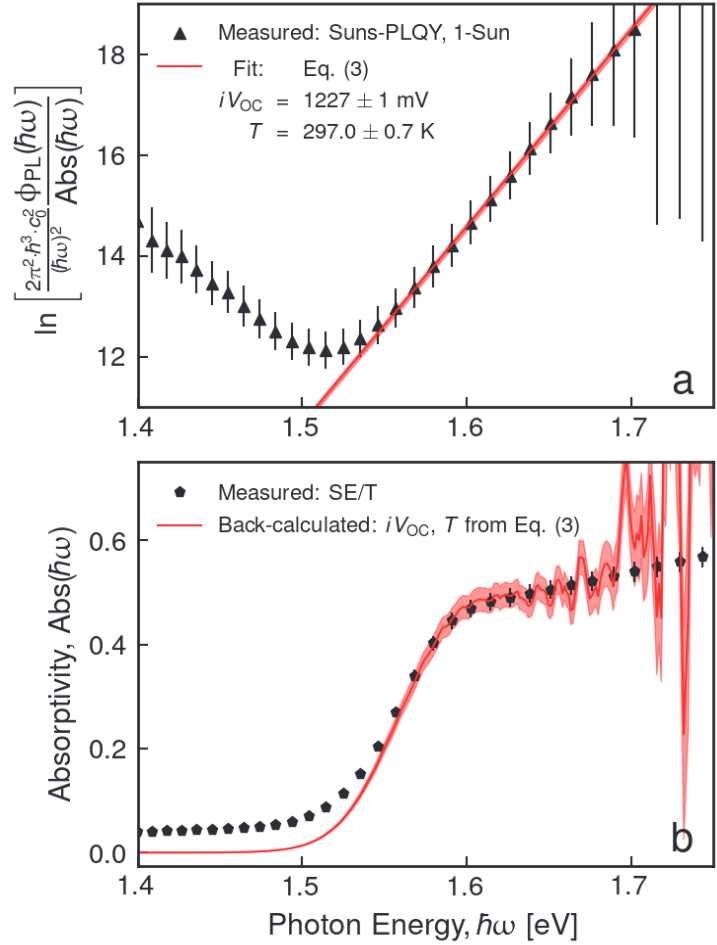

**Figure 2** (a) A linear transform of the 1-Sun $\phi_{PL}(\hbar\omega)$, according to Equation (3) (black triangles) and linear curve fit (solid red line) used to extract $iV_{OC}$ and $T$. The error bars on the black triangles represent the measurement uncertainty. The shaded red regions represent the effect of the standard errors of $iV_{OC}$ and $T$ on the fit. (b) Reference spectral absorptivity (black pentagons) and back-calculated spectral absorptivity from 1-Sun $\phi_{PL}(\hbar\omega)$ (solid red line). The error bars on the measured spectral absorptivity represent the standard deviation of the absorptivity measured from either side of the PTF/glass stack. The shaded red regions represent the propagated uncertainty in the measured $\phi_{PL}(\hbar\omega)$, and the curve-fitted $iV_{OC}$ and $T$.

The $iV_{OC,rad}$ values are determined using Equation (3) together with the determined $iV_{OC}$ and PLQY. **Figure 3** illustrates the Suns-PLQY data represented as dark $iJ$-$V$ curves, where black triangles (circles) represent the dark (radiative) $iJ$-$V$ curve. A $\Delta iV_{OC}$ of 30 mV at 1-Sun, caused by NRBR, is indicated by the dashed horizontal line. By fitting the dark radiative $iJ$-$V$ curve, we obtain a $J_{0,rad}$ value of (12.2 ± 0.1) yocto-A.cm$^{-2}$ (1 yocto = 10$^{-24}$) using Equation (6). Subsequently, utilising this $J_{0,rad}$ value and Equation (8), we determine a $J_{0,nr}$ value of (36.5 ± 0.5) yocto-A.cm$^{-2}$ from the fit to the dark $iJ$-$V$ curve.

It is important to note that the fitting region for the dark $iJ$-$V$ curve is limited to $iV > 1,200$ mV. Below this threshold, the curve deviates from Equation (8), likely due to the contribution of other recombination mechanisms such as bulk Shockley-Read-Hall (SRH) [30]. The dark $iJ$-$V$ curve fits for the thick and thin PTFs are in Section S4 of SI.

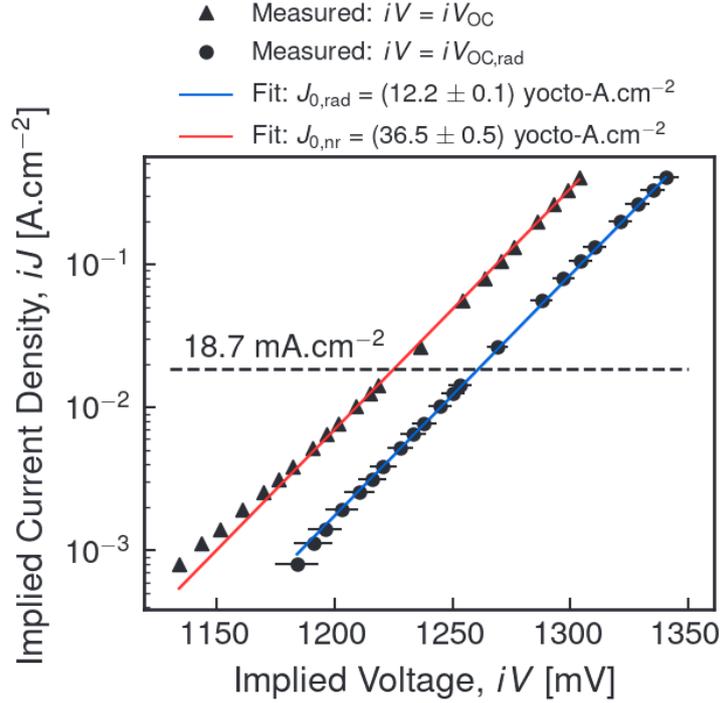

**Figure 3** Dark *iJ-V* curves for intermediate PTF. The red (blue) curve is a curve fit to the dark (radiative) *iJ-V* curve using Equation (8) [Equation (6)]. The dashed line represents the 1-Sun $J_L$, equal to 18.7 mA.cm$^{-2}$. The horizontal distance between the intersection of the dark *iJ-V* curves with this dashed line represents the 1-Sun $\Delta iV_{OC}$.

**Table 2** lists the $J_{0,nr}$ values for PTFs of varying thickness, indicating a slight increase in $J_{0,nr}$ with increasing thickness (less than 33%, see Section S6 of the SI). If NRBR originates only from the bulk, one would anticipate a nearly tripled $J_{0,nr}$ when transitioning from thin to thick PTFs (469/160 nm ≈ 2.93). Conversely, if NRBR arises from the surfaces, $J_{0,nr}$ should be independent of the thickness. The observed modest increase in $J_{0,nr}$ with increasing thickness suggests the presence of a significant surface component combined with a relatively smaller bulk component of $J_{0,nr}$.

While a bulk $J_{0,nr}$ component could be attributed to SRH recombination [30] from shallow bulk defects, we find that the required $N_t$ is implausibly large (see Section S5 of SI). A simpler explanation is relatively minor differences in the SRH recombination parameters [31] among the PTFs of different thicknesses. For instance, a 33% variation in the interface defect density, $N_{it}$, with the thickness could account for the observed difference. Therefore, the most plausible cause of NRBR are defects at the surfaces, and in this context, the $J_{0,nr}$ parameter is equivalent to $J_{0,s}$ (surface recombination). It should be noted that $J_{0,s}$ represents the cumulative value across both PTF/air and PTF/glass surfaces.

It is important to note that if NRBR originated from the grain boundaries, the opposite trend would be expected between $J_{0,nr}$ and the grain boundary diameter, $D_{gb}$ (also listed in **Table 2,** see Section S6 of the SI) [32], since the grain boundary area per unit volume of the film is inversely proportional to $D_{gb}$ (see Section S6 of SI). Hence, we rule out grain boundaries as the possible spatial origin of $J_{0,nonrad}$.

Typically, surface recombination in PTFs has been specified in terms of the surface recombination velocity, SRV (cm.s$^{-1}$) [33], [34]. SRV values are usually specified as injection-independent values, meaning the surface recombination rate is linear with respect to $\Delta n$ (monomolecular recombination). However, when the surfaces are highly charged, corresponding to strong accumulation or strong inversion, the electron and hole densities at the surfaces are highly asymmetric. This leads to an injection-dependent SRV more aptly parametrised by the *surface* saturation current parameter, $J_{0,s}$ [31] and surface recombination proportional to the electron-hole product (bimolecular recombination). We note that $J_{0,s}$ is a well-established parameter for crystalline silicon photovoltaics [35], routinely used to assess the surface passivation quality [31], [36].

**Table 2** $J_{0,nr}$ values obtained from Equations (6) and (8). The asterisks indicate additional measurements on separate locations. $W$ and $D_{gb}$ values are included for completeness. The average uncertainty is calculated using $\sqrt{\langle J_{0,nr}^2 + \Delta J_{0,nr}^2 \rangle - \langle J_{0,nr} \rangle^2}$, where $\Delta J_{0,nr}$ is the individual uncertainty of each $J_{0,nr}$ measurement and the brackets indicate the mean value.

| Thickness | $W$ [nm] | $D_{gb}$ [nm] | $J_{0,nr}$ [yocto-A.cm$^{-2}$] |
|---|---|---|---|
| Thick | 469.3 ± 4.0 | 203 ± 64 | 40.6 ± 1.4 |
| Intermediate | 262.2 ± 2.0 | 170 ± 52 | 36.5 ± 0.5 |
| Intermediate* | | | 36.2 ± 0.5 |
| Thin | 159.5 ± 5.5 | 151 ± 55 | 31.2 ± 0.5 |
| Thin* | | | 33.4 ± 0.4 |
| Average | n/A | | 35.6 ± 10.6 |

Regarding the origin of $J_{0,s}$, we propose two possible mechanisms from the framework of SRH recombination [30]. Both assume a single energy-level surface defect and involve a significant interface band-bending/photovoltage, $\psi_s$, leading to asymmetric electron and hole densities at the boundary [31]. The first case is near band-edge donor (acceptor) states close to the conduction (valence) band-edge:

$$J_{0,s} = q \cdot c_x \cdot N_{it} \cdot n_i \cdot \exp\left[-\frac{E_t - E_i}{k_B T}\right] \quad (10)$$

$c_x$ is the capture coefficient with $x = p$ ($x = n$) for a donor (acceptor), $N_{it}$ is the interface state density (cm$^{-2}$) and ($E_t - E_i$) is the defect energy level referenced to the intrinsic level. We determine that Equation (10) is a reasonable approximation if the defect energy-level is within 50 meV of the valence (conduction) band-edge, corresponding to excess carrier densities of less than $10^{17}$ cm$^{-3}$. This mechanism was previously proposed by Brenes *et al* [18]. The second case is a mid-gap defect accompanied by a large surface charge, $Q_s$ [31]:

$$J_{0,s} = \frac{q \cdot c_x \cdot N_{it} \cdot 2k_B T \cdot \varepsilon \cdot n_i^2}{Q_s^2} \quad (11)$$

where $\varepsilon$ is the permittivity. $Q_s$ could be caused by the accumulation of mobile, charged ions near the surfaces [37]. We performed surface photovoltage mapping on a Br17 sample with in-situ LS, revealing a positive $\psi_s$ of several hundred meV. This indicates the presence of donor states at the surface, further reinforcing the hypothesis of a significant $J_{0,s}$ (see Section S8 of the SI). To distinguish which of these two mechanisms is relevant, the temperature-dependence of $J_{0,s}$ can be measured from temperature-dependent Suns-PLQY. Ignoring the temperature dependencies of $c_n$, $Q_s$, and $\varepsilon$, which are non-trivial, Equations (10) and (11) have distinct temperature dependencies of $T^{3/2} \cdot \exp(-[\underline{E_g}(T) + 2 \cdot (E_t - E_i)]/2k_BT)$ and $T^4 \cdot \exp(-E_g(T)/k_BT)$, respectively. Note that $E_g$ is the bandgap energy.

## 5. Conclusions

In this study, we have successfully showcased a straightforward analysis approach utilising Suns-PLQY measurements to quantify non-radiative bimolecular recombination in semiconductor thin films, as well as identify its spatial origin. Notably, we demonstrated that NRBR in the studied PTFs arises from defects located at the surfaces, which can be effectively described by the $J_{0,s}$ parameter. We determined an average $J_{0,s}$ value of (35.6 ± 10.6) yocto-A.cm$^{-2}$ across thicknesses ranging from 160 nm to 470 nm. We propose either band-edge surface defects or mid-gap surface defects coupled with large surface charge as plausible origins of $J_{0,s}$, and suggest using temperature-dependent Suns-PLQY to determine the exact origin of $J_{0,s}$. The application of this simple analysis technique holds potential for the characterisation of $J_{0,s}$ in perovskite compositions and other semiconductor materials.

## Acknowledgements

This Project received funding from the Australian Renewable Energy Agency (ARENA) as part of the TRAC Program (Grant 2022/TRAC001). The views expressed herein are not necessarily the views of the Australian Government, and the Australian Government does not accept responsibility for any information or advice contained herein.

# References

[1] M. Stolterfoht *et al.*, 'How To quantify the efficiency potential of neat Perovskite films: Perovskite semiconductors with an implied efficiency exceeding 28%', *Advanced Materials*, vol. 32, no. 17, p. 2000080, 2020, doi: https://doi.org/10.1002/adma.202000080.

[2] C. J. Hages *et al.*, 'Identifying the real minority carrier lifetime in nonideal semiconductors: A case study of Kesterite materials', *Advanced Energy Materials*, vol. 7, no. 18, p. 1700167, 2017, doi: 10.1002/aenm.201700167.

[3] J. C. de Mello, H. F. Wittmann, and R. H. Friend, 'An improved experimental determination of external photoluminescence quantum efficiency', *Advanced Materials*, vol. 9, no. 3, pp. 230–232, 1997, doi: 10.1002/adma.19970090308.

[4] J. K. Katahara and H. W. Hillhouse, 'Quasi-Fermi level splitting and sub-bandgap absorptivity from semiconductor photoluminescence', *Journal of Applied Physics*, vol. 116, no. 17, p. 173504, 2014, doi: 10.1063/1.4898346.

[5] I. L. Braly *et al.*, 'Hybrid perovskite films approaching the radiative limit with over 90% photoluminescence quantum efficiency', *Nature Photonics*, vol. 12, no. 6, pp. 355–361, 2018, doi: 10.1038/s41566-018-0154-z.

[6] K. Frohna *et al.*, 'Nanoscale chemical heterogeneity dominates the optoelectronic response of alloyed perovskite solar cells', *Nature Nanotechnology*, vol. 17, no. 2, pp. 190–196, 2022, doi: 10.1038/s41565-021-01019-7.

[7] P. Wurfel, 'The chemical potential of radiation', *Journal of Physics C: Solid State Physics*, vol. 15, no. 18, pp. 3967–3985, 1982, doi: 10.1088/0022-3719/15/18/012.

[8] U. Rau, 'Reciprocity relation between photovoltaic quantum efficiency and electroluminescent emission of solar cells', *Physical Review B*, vol. 76, no. 8, p. 085303, 2007, doi: 10.1103/PhysRevB.76.085303.

[9] P. Caprioglio *et al.*, 'On the origin of the ideality factor in perovskite solar cells', *Advanced Energy Materials*, p. 2000502, 2020, doi: 10.1002/aenm.202000502.

[10] V. Sarritzu *et al.*, 'Optical determination of Shockley-Read-Hall and interface recombination currents in hybrid perovskites', *Scientific Reports*, vol. 7, 2017, doi: 10.1038/srep44629.

[11] J. M. Richter *et al.*, 'Enhancing photoluminescence yields in lead halide perovskites by photon recycling and light out-coupling', *Nature Communications*, vol. 7, no. 1, p. 13941, 2016, doi: 10.1038/ncomms13941.

[12] F. Staub, T. Kirchartz, K. Bittkau, and U. Rau, 'Manipulating the net radiative recombination rate in lead halide perovskite films by modification of light outcoupling', *The Journal of Physical Chemistry Letters*, vol. 8, no. 20, pp. 5084–5090, 2017, doi: 10.1021/acs.jpclett.7b02224.

[13] C. L. Davies *et al.*, 'Bimolecular recombination in methylammonium lead triiodide perovskite is an inverse absorption process', *Nature Communications*, vol. 9, no. 1, p. 293, 2018, doi: 10.1038/s41467-017-02670-2.

[14] A. Simbula *et al.*, 'Direct measurement of radiative decay rates in metal halide perovskites', *Energy & Environmental Science*, vol. 15, no. 3, pp. 1211–1221, 2022, doi: 10.1039/D1EE03426J.

[15] A. R. Bowman, S. Macpherson, A. Abfalterer, K. Frohna, S. Nagane, and S. D. Stranks, 'Extracting decay-rate ratios from photoluminescence quantum efficiency measurements in optoelectronic semiconductors', *Physical Review Applied*, vol. 17, no. 4, p. 044026, 2022, doi: 10.1103/PhysRevApplied.17.044026.

[16] F. Staub, U. Rau, and T. Kirchartz, 'Statistics of the Auger recombination of electrons and holes via defect levels in the band gap—Application to lead-halide perovskites', *ACS Omega*, vol. 3, no. 7, pp. 8009–8016, 2018, doi: 10.1021/acsomega.8b00962.

[17] A. Kiligaridis *et al.*, 'Are Shockley-Read-Hall and ABC models valid for lead halide perovskites?', *Nature Communications*, vol. 12, no. 1, p. 3329, 2021, doi: 10.1038/s41467-021-23275-w.

[18] R. Brenes *et al.*, 'Metal halide perovskite polycrystalline films exhibiting properties of single crystals', *Joule*, vol. 1, no. 1, pp. 155–167, 2017, doi: 10.1016/j.joule.2017.08.006.

[19] A. Hangleiter, 'Nonradiative recombination via deep impurity levels in silicon: Experiment', *Physical Review B*, vol. 35, no. 17, pp. 9149–9161, Jun. 1987, doi: 10.1103/PhysRevB.35.9149.

[20] G. F. Burkhard, E. T. Hoke, and M. Group, 'Transfer Matrix Optical Modeling'.

[21] P. Fassl *et al.*, 'Revealing the internal luminescence quantum efficiency of perovskite films via accurate quantification of photon recycling', *Matter*, vol. 4, no. 4, pp. 1391–1412, 2021, doi: 10.1016/j.matt.2021.01.019.

[22] G. Lasher and F. Stern, 'Spontaneous and stimulated recombination radiation in semiconductors', *Physical Review*, vol. 133, no. 2A, pp. A553–A563, 1964, doi: 10.1103/PhysRev.133.A553.

[23] G. Rey *et al.*, 'Absorption coefficient of a semiconductor thin film from photoluminescence', *Physical Review Applied*, vol. 9, no. 6, p. 064008, 2018, doi: 10.1103/PhysRevApplied.9.064008.

[24] G. El-Hajje *et al.*, 'Quantification of spatial inhomogeneity in perovskite solar cells by hyperspectral luminescence imaging', *Energy & Environmental Science*, vol. 9, no. 7, pp. 2286–2294, 2016, doi: 10.1039/C6EE00462H.

[25] M. Tebyetekerwa *et al.*, 'Quantifying quasi-Fermi level splitting and mapping its heterogeneity in atomically thin transition metal dichalcogenides', *Advanced Materials*, vol. 31, no. 25, p. 1900522, 2019, doi: 10.1002/adma.201900522.

[26] A. Cuevas, 'The recombination parameter $J_0$', *Energy Procedia*, vol. 55, pp. 53–62, 2014, doi: 10.1016/j.egypro.2014.08.073.

[27] P. P. Altermatt, A. Schenk, F. Geelhaar, and G. Heiser, 'Reassessment of the intrinsic carrier density in crystalline silicon in view of band-gap narrowing', *Journal of Applied Physics*, vol. 93, no. 3, pp. 1598–1604, 2003, doi: 10.1063/1.1529297.

[28] D. J. Slotcavage, H. I. Karunadasa, and M. D. McGehee, 'Light-induced phase segregation in halide-perovskite absorbers', *ACS Energy Letters*, vol. 1, no. 6, pp. 1199–1205, 2016, doi: 10.1021/acsenergylett.6b00495.

[29] T. Kirchartz, J. A. Márquez, M. Stolterfoht, and T. Unold, 'Photoluminescence-based characterization of halide Perovskites for photovoltaics', *Advanced Energy Materials*, p. 1904134, 2020, doi: 10.1002/aenm.201904134.

[30] W. Shockley and W. T. Read, 'Statistics of the recombinations of holes and electrons', *Physical Review*, vol. 87, no. 5, pp. 835–842, 1952, doi: 10.1103/PhysRev.87.835.

[31] K. R. McIntosh and L. E. Black, 'On effective surface recombination parameters', *Journal of Applied Physics*, vol. 116, no. 1, p. 014503, 2014, doi: 10.1063/1.4886595.

[32] M. Yang *et al.*, 'Do grain boundaries dominate non-radiative recombination in $CH_3NH_3PbI_3$ perovskite thin films?', *Physical Chemistry Chemical Physics*, vol. 19, no. 7, pp. 5043–5050, 2017, doi: 10.1039/C6CP08770A.

[33] J. Wang, W. Fu, S. Jariwala, I. Sinha, A. K.-Y. Jen, and D. S. Ginger, 'Reducing Surface Recombination Velocities at the Electrical Contacts Will Improve Perovskite Photovoltaics', *ACS Energy Letters*, vol. 4, no. 1, pp. 222–227, 2019, doi: 10.1021/acsenergylett.8b02058.

[34] Y. Yang *et al.*, 'Top and bottom surfaces limit carrier lifetime in lead iodide perovskite films', *Nature Energy*, vol. 2, no. 2, p. 16207, 2017, doi: 10.1038/nenergy.2016.207.

[35] D. E. Kane and R. M. Swanson, 'Measurement of the emitter saturation current by a contactless photoconductivity decay method', in *18th IEEE Photovoltaic Specialist Conference*, 1985, pp. 578–583.

[36] T. Niewelt *et al.*, 'Reassessment of the intrinsic bulk recombination in crystalline silicon', *Solar Energy Materials and Solar Cells*, vol. 235, p. 111467, 2022, doi: 10.1016/j.solmat.2021.111467.

[37] D. S. Gets, G. A. Verkhogliadov, E. Y. Danilovskiy, A. I. Baranov, S. V. Makarov, and A. A. Zakhidov, 'Dipolar cation accumulation at the interfaces of perovskite light-emitting solar cells', *J. Mater. Chem. C*, vol. 8, no. 47, pp. 16992–16999, 2020, doi: 10.1039/D0TC02654A.

[38] Y. Jiang, M. A. Green, R. Sheng, and A. Ho-Baillie, 'Room temperature optical properties of organic–inorganic lead halide perovskites', *Solar Energy Materials and Solar Cells*, vol. 137, pp. 253–257, 2015, doi: 10.1016/j.solmat.2015.02.017.

[39] Y. Jiang, A. M. Soufiani, A. Gentle, F. Huang, A. Ho-Baillie, and M. A. Green, 'Temperature dependent optical properties of $CH_3NH_3PbI_3$ perovskite by spectroscopic ellipsometry', *Applied Physics Letters*, vol. 108, no. 6, p. 061905, 2016, doi: 10.1063/1.4941710.

[40] F. Staub *et al.*, 'Beyond bulk lifetimes: Insights into lead halide Perovskite films from time-resolved photoluminescence', *Physical Review Applied*, vol. 6, no. 4, p. 044017, 2016, doi: 10.1103/PhysRevApplied.6.044017.

[41] F. E. Rougieux, C. Sun, and D. Macdonald, 'Determining the charge states and capture mechanisms of defects in silicon through accurate recombination analyses: A review', *Solar Energy Materials and Solar Cells*, vol. 187, pp. 263–272, 2018, doi: 10.1016/j.solmat.2018.07.029.

[42] T. Kirchartz, 'High open-circuit voltages in lead-halide perovskite solar cells: experiment, theory and open questions', *Philosophical Transactions of the Royal Society A: Mathematical, Physical and Engineering Sciences*, vol. 377, no. 2152, p. 20180286, Aug. 2019, doi: 10.1098/rsta.2018.0286.

[43] E. Choi *et al.*, 'Exploration of sub-bandgap states in 2D halide perovskite single-crystal photodetector', *npj 2D Mater Appl*, vol. 6, no. 1, p. 43, 2022, doi: 10.1038/s41699-022-00317-5.